
\magnification=1200
\pageno=0\tolerance=10000\hfuzz=5pt
\line{\hfill CERN-TH.7189/94}
\line{\hfill   DFTT 02/94}
\line{\hfill March 1994}
\vskip 10pt
\centerline{\bf ANOMALOUS EVOLUTION OF}
\centerline{\bf NONSINGLET NUCLEON STRUCTURE FUNCTIONS}
\vskip 36pt
\centerline{Richard  D.~Ball,$^{a}$\footnote*{On leave from a Royal
Society University Research Fellowship.} Vincenzo Barone,$^{b}$
Stefano Forte$^a$\footnote\dag{On leave from INFN, Sezione di
Torino, Italy} and
Marco Genovese$^{b}$}
\vskip 12pt
\centerline{\it {}$^a$ Theory Division, CERN}
\centerline{\it CH-1211, Gen\`eve 23, Switzerland}
\vskip 10pt
\centerline{\it and}
\vskip 10pt
\centerline{\it {}$^b$Dipartimento di Fisica Teorica dell'Universit\`a
                       and INFN}
\centerline{\it via P.~Giuria 1, I-10125 Torino, Italy}
\vskip .55in
{\narrower\baselineskip 10pt
\centerline{\bf Abstract}
\medskip
\noindent
We calculate the scale dependence of nonsinglet nucleon
structure functions. Due
to anomalous axial symmetry breaking a large flavour asymmetry of
the quark--antiquark sea is generated nonperturbatively. This produces a
strong scale dependence of the nonsinglet structure function in an
intermediate range of $Q^2$. Evolving nonperturbatively a pure valence
distribution from an infrared scale we can thus compute
$F_2^p-F_2^n$ as measured by the NMC, and give detailed predictions for
its $Q^2$ dependence at fixed $x$.
We also compare our results with Drell--Yan data.
}
\vskip 0.5in
\centerline{Submitted to: {\it Physics Letters B}}
\vskip 0.5in
\line{CERN-TH.7189/94 \hfill}
\line{DFTT 02/94\hfill}
\line{February 1994\hfill}
\vfill\eject

\input harvmac

\def\ie{{\it i.e.}}

\def\etal{{\it et al.}}

\def\p2inf{\mathrel{\mathop{\sim}\limits_{\scriptscriptstyle
{p^2 \rightarrow \infty }}}}
\def\k2inf{\mathrel{\mathop{\sim}\limits_{\scriptscriptstyle
{k^2 \rightarrow \infty }}}}
\def\x2inf{\mathrel{\mathop{\sim}\limits_{\scriptscriptstyle
{x \rightarrow \infty }}}}
\def\Lam2inf{\mathrel{\mathop{\sim}\limits_{\scriptscriptstyle
{\Lambda \rightarrow \infty }}}}
\def\Q2inf{\mathrel{\mathop{\sim}\limits_{\scriptscriptstyle
{Q^2 \rightarrow \infty }}}}

\def\frac#1#2{{{#1}\over {#2}}}
\def\smallfrac#1#2{\hbox{$\frac{#1}{#2}$}}

\def\third{\smallfrac{1}{3}}

\catcode`@=11 
\def\slash#1{\mathord{\mathpalette\c@ncel#1}}
 \def\c@ncel#1#2{\ooalign{$\hfil#1\mkern1mu/\hfil$\crcr$#1#2$}}
\def\lsim{\mathrel{\mathpalette\@versim<}}
\def\gsim{\mathrel{\mathpalette\@versim>}}
 \def\@versim#1#2{\lower0.2ex\vbox{\baselineskip\z@skip\lineskip\z@skip
       \lineskiplimit\z@\ialign{$\m@th#1\hfil##$\crcr#2\crcr\sim\crcr}}}
\catcode`@=12 

\def\PR{{\it Phys.~Rev.~}}
\def\PRL{{\it Phys.~Rev.~Lett.~}}
\def\NP{{\it Nucl.~Phys.~}}
\def\PL{{\it Phys.~Lett.~}}
\def\PRep{{\it Phys.~Rep.~}}
\def\AP{{\it Ann.~Phys.~}}

\def\ZP{{\it Zeit.~Phys.~}}

\def\IJMP{{\it Int.~Jour.~Mod.~Phys.~}}

\def\vol#1{{\bf #1}}
\def\vyp#1#2#3{\vol{#1} (#2) #3}
\def\vpy#1#2#3{\vol{#1} (#3) #2}


The detailed description of logarithmic scaling violations is perhaps the
most successful result of perturbative QCD, particularly in the context of
deep inelastic scattering, where the scale dependence
of the nucleon structure functions can be accurately computed
perturbatively either through the operator product expansion or
equivalently by solving the Altarelli--Parisi equations. Recent improvements
in experimental accuracy are now opening up the possibility of
measuring subleading perturbative corrections, as well as corrections
to perturbation theory as expressed by higher twist effects. However,
it has been recently suggested~\ref\balfor{R.~D.~ Ball and S.~Forte, Oxford
preprint OUTP-93-18P and Torino preprint DFTT 9/93 (1993).} that in an
intermediate region of $Q^2\sim 1$~GeV$^2$  the scale dependence of
the nonsinglet nucleon structure function may be qualitatively rather
different from that predicted by purely perturbative QCD, due to
nonperturbative effects related to the anomalous breaking of axial
U(1) symmetry.

In ref.\balfor\ it was shown that recent data on the first moment
of the nonsinglet nucleon structure function~\nref\nmca{P.~Amaudruz \etal,
\PRL\vpy{66}{560}{1991}.}\nref\nmcb{P.~Amaudruz \etal,
CERN preprint CERN-PPE/93-117 (1993).}\refs{\nmca,\nmcb},
which seem to contradict standard
assumptions on the symmetry of the light quark sea in the nucleon, can
be understood in a natural way if nonperturbative
scaling violations are taken into account by suitable modification
of the Altarelli--Parisi equations. These scaling volations are
relatively large in the nonsinglet channel due to the fact that
the ${\rm U}(N_f)$ symmetry of the spectrum of
light pseudoscalar bound states (the Goldstone bosons of spontaneous
chiral symmetry breakdown) is broken explicitly down to ${\rm
SU}(N_f)$ by the axial anomaly and fluctuations of topological charge
in the QCD vacuum. If this (anomalous) symmetry breaking
effect, which is essentially nonperturbative~\ref\uone{G.~'t~Hooft,
\PRep\vpy{142}{357}{1986} and ref.
therein.}, is taken into account in the
evolution equations, then the first moments of nonsinglet charge conjugation
even structure functions, which perturbatively are almost scale
independent, acquire a strong scale dependence
for $Q^2$ up to a few GeV$^2$, due to the generation of a
flavour asymmetric component of the quark--antiquark sea.

Here we will investigate  the
consequences of such a modification of the evolution equations on the
$Q^2$ dependence of the full $x$-dependent
nonsinglet, charge conjugation even
structure function \hbox{$[F_2^p(x,Q^2)-F_2^n(x,Q^2)]$.} We will show that
the shape of this structure function is also significantly affected
by the production of the asymmetric sea induced by the
nonperturbative effects mentioned above, in a way which is
experimentally verifiable independently of the effect on the first
moment. We will show further that if one
attempts to compute the shape of the nonsinglet structure function
when $Q^2$ is of the order of several GeV$^2$ by evolving some valence
quark distribution from an infrared scale, then the inclusion of
our nonperturbative effects leads to a qualitative improvement in
the agreement with the NMC data, which is independent of the details
of the initial valence quark distribution. Finally, we will show that
our results are also in agreement with available data from the Drell--Yan
process on the $x$-dependence of the ratio of $\bar u$ and $\bar d$
sea quark distributions.

\medskip

Let us first summarize briefly the formalism developed in ref.\balfor\ to
include anomalous symmetry breaking in the nonsinglet Altarelli--Parisi
evolution equations.  We will concentrate on the charge  conjugation
even sector, \ie\ on the combination
\eqn\qplus{q^+_i(x)\equiv q_i(x)+\bar q_i(x)}
of the quark and antiquark distributions of flavour $i$. These
are directly measurable in deep inelastic scattering, since in the
parton model the structure function for unpolarized electroproduction
$F_2$ is given (to all orders, in the parton
scheme~\ref\alta{G.~Altarelli, \PRep\vpy{81}{1}{1982}.}) by
\eqn\fandq
{F_2(x)\equiv x\sum_ie^2_iq^+_i(x),}
where $e_i$ are the electric charges of the quarks.

Now, in perturbative QCD the evolution equation for any nonsinglet
combination $q^+(x)$ of the distributions $q^+_i(x)$ (for example
$q^+=u^+(x)-d^+(x)$) is given by \alta
\eqn\evolqp
{{d\over dt}{q^+}=
\left({\cal Q}_{qq}+ {\cal Q}_{q\bar q}\right)\otimes {q^+} ,}
where
\eqn\qdef
{{\cal Q}_{qq}\equiv {\cal P}_{qq}^D-{\cal P}_{qq}^{ND},}
and ${\cal P}^D$ (${\cal P}^{ND}$) is any quark--quark splitting function
${\cal P}_{q_iq_j}$ which is diagonal (nondiagonal) in flavour, \ie,
such that $i=j$ ($i\not =j$), and analogously for $Q_{q\bar q}$.
At one loop, all the splitting functions ${\cal P}^{ND}$ and
${\cal P}_{q\bar q}$  vanish, hence the evolution of $q^+$ is
driven entirely by the quark--quark splitting function
${\cal P}_{qq}^{D}$: the
quarks lose momentum by radiation of gluons. This leaves the first
moment of $q^+$ unchanged (by charge conservation), whereas all
higher moments necessarily decrease. Thus at one loop
a nonsinglet
quark-antiquark sea cannot be generated radiatively:
if the sea vanishes at some scale, it will vanish at all scales, and
only the valence will evolve.
At two loops all the splitting functions are nonzero, but the
two loop contribution to the evolution of $q^+$ eqn.\evolqp\
is extremely small, because ${\cal P}^D$ and ${\cal P}^{ND}$ only differ
due to antisymmetrization of the sea quark pairs in the final state~\ref\rs{
D.~A.~Ross and C.~T.~Sachrajda, \NP\vpy{B149}{497}{1979}.}.
Indeed, the U(N$_f$) flavour symmetry of the QCD Lagrangian ensures that
perturbatively no nonsinglet component of the sea can be generated,
modulo the tiny final state antisymmetrization effect.

The nonperturbative breaking of the U(N$_f$) symmetry due to the axial
anomaly can be incorporated into the evolution equations  by
considering the contribution to $Q^2$ evolution from diagrams where
the quark couples directly to a $q\bar q$ bound state. These can be
viewed as being generated dynamically from the diagram which
provides the two loop contribution to the usual splitting function
${\cal P}_{q_iq_j}$, but where the emitted (and unobserved)
quark--antiquark pair is allowed to bind nonperturbatively into a
meson state. A full, self--consistent set of coupled evolution
equations for quarks and mesonic bound states in the nonsinglet
channel can then be written down \balfor;
it turns out that the evolution equation for $q^+$ retains the
form of eqn.\qplus, but now the splitting functions $\cal P$ receive
contributions due to (one loop) bound state emission.
The splitting functions can thus be computed explicitly from the
respective particles' emission cross section, according to
\eqn\splitX
{\left[{\cal P}_{q_iq_j}(x;t)\right]_X=
\frac{d}{dt}\sigma^{\gamma^*X}_{q_iq_j}(x;t),}
where $\sigma^{\gamma^*X}_{q_iq_j} (x;t)$ is the total cross section for
absorption of a virtual photon $\gamma^*$ and
emission of the state $X$, expressed in terms of
the usual scaling variables and integrated over all $k_\perp$.

Because the bound states carry flavour, ${\cal P}^{ND}$ is nonvanishing
already at leading order in the quark--meson coupling. Furthermore,
the combination of spontaneous  chiral symmetry breaking
SU(N$_f$)$\times$SU(N$_f$)$\to$SU(N$_f$) and the explicit anomalous
breaking of the axial U(1), resulting in the relatively large
$\pi$-$\eta^\prime$ mass splitting, leads kinematically to a
substantial difference between the splitting functions ${\cal P}^D$
and ${\cal P}^{ND}$, and thus to the generation of a sizable
nonsinglet component in the quark--antiquark sea. As the
breaking of U(N$_f$) chiral symmetry is only manifested in the
pseudoscalar meson channel, these are actually
the only bound states which contribute significantly to the
nonsinglet evolution; for other multiplets the splitting functions
${\cal P}^D$ and ${\cal P}^{ND}$ are almost equal, and are in any case
suppressed kinematically (by at least  an order of magnitude) due to the
relatively large masses of the emitted bound states.

The full splitting function \qdef\ is thus determined according to
eqn.\splitX\ in terms of the set of pseudoscalar octet meson emission
cross sections, which, due to the extended nature of the bound states,
depend on a quark--meson vertex function. This may be decomposed as a
sum of four independent couplings  (a pseudoscalar, two axial, and a
tensor), yielding four independent (scalar) vertex functions.
However, once the cross section is computed\balfor,
it turns out to depend only on
a pseudoscalar
vertex function  $\varphi$ and an axial vertex function
$\tilde\varphi$; the former dominates  for
intermediate values of $Q^2$ (up to  a few GeV$^2$),  while
the latter controls  the
large $Q^2$ tail. The pseudoscalar vertex function, which,
consistent with known asymptotic behaviour, can be
taken to
have the simple dipole form
\eqn\ffp
{\varphi(p^2)=\frac{m_d}{f_\pi}\frac{\Lambda^2+m_d^2}{\Lambda^2+p^2},}
can be further constrained by using the chiral Ward identity to
express the constituent quark mass $m_d$ in terms of
the parameter $\Lambda$; this restricts the allowed range of values of
$\Lambda$ to be $0.4$~GeV~$\lsim\Lambda\lsim 0.8$~GeV.
The axial vertex function may be assumed to have the similar form
\eqn\ffa
{\tilde\varphi(p^2)=\frac{g_\pi}{f_\pi}\frac{\tilde\Lambda^2+m_d^2}
{\tilde\Lambda^2+p^2},}
with $\tilde \Lambda\sim\Lambda$ while the parameter $g_\pi$
is likely to be rather small,   say
$g_\pi\lsim{1\over2}$.

An explicit computation of the various cross sections \balfor\ shows
that even though the vertex function \ffp\ is soft
(\ie\ $\Lambda$ is rather small)
the contribution to the anomalous dimensions calculated from
eqn.\splitX\ remains sizable for values of $Q^2$ up to 5--10~GeV$^2$
because there exists a region of phase space (at small $x$ in the
$t$-channel and at large $x$ in the $s$-channel) where the vertex function is
pointlike for arbitrarily large values of $Q^2$. Furthermore it is
precisely in this dominant region that the emitted bound state and
quark are quasi--collinear, which
justifies the use of a parton picture, and the Altarelli--Parisi
evolution equations. However, at both very small
$Q^2\sim 0.05$~GeV$^2$, and at large $Q^2 > 10$~GeV$^2$, all the
anomalous dimensions flatten, ensuring both a
smooth connection to a valence quark picture in the infrared and to
the usual perturbative behaviour in the ultraviolet.

\medskip

We wish now to solve the evolution equation \evolqp\ for the
nonsinglet combination of structure functions
\eqn\nsstrf
{{1\over x}\big(F_2^p(x)-F_2^n(x)\big)\equiv\third q^{+}(x)
\equiv\third\left[q^v(x)+2\bar q(x)\right],}
using the nonperturbative splitting function due to
bound state emission determined from eqn.\splitX\
(and given explicitly in the appendix of ref.\balfor).
If in eqn.\nsstrf\ we assume exact isospin symmetry, $q^+=u^+-d^+$
is just the nonsinglet combination of quark distributions
eqn.\qplus\ with isospin one; in the last step we have introduced
for convenience valence quark distributions defined as
\hbox{$q^v(x)=q(x)-\bar q(x)$.}
Whereas the computation of the evolution of the first moment of
$q^+$ (as performed in ref.\balfor) required only the
nonperturbative splitting function which governs the
generation of an asymmetric sea component due to bound state emission
(the perturbative evolution of the first moment being
negligible), the determination of the evolution of the full
structure function requires in addition the contribution to
the splitting function which determines the evolution of the
valence distribution due to gluon emission. Furthermore, whereas
the initial condition for the evolution of the first moment is
very simple --- it is just the flavour asymmetry of the quark sea
at the  `constituent quark' scale, which can be taken to be zero ---
that for the full distribution is nontrivial: even assuming that
there is initially no asymmetry in the sea one still needs to
know the initial shape of the nonsinglet valence distribution, eqn.\nsstrf.

In order to actually perform the computation, we will assume that bound
state emission contributes significantly only to the generation of an
asymmetric sea. This
is in keeping with the observation that the anomalous dimensions for
bound state emission are generally much smaller than the usual perturbative
ones, and are indeed only significant at all because of
the nonperturbative breaking
of the axial symmetry. The gluon emission responsible for evolution of
the valence distribution, and the bound state emission responsible for
the nonsinglet sea evolution, are then to be regarded as  truly distinct
mechanisms. The full splitting functions to be used in
eqns~\evolqp\ and \qdef\ are thus given by
\eqn\splf{
{\cal P}_{qq}^{ND}=\left[{\cal P}_{qq}^{\Pi}\right]^{ND},\qquad
{\cal P}_{qq}^D=\left[{\cal P}_{qq}^\Pi\right]^{D}+\left[{\cal
P}_{qq}^g
\right]^{D},}
where ${\cal P}^\Pi$ are the nonperturbative splitting functions due
to meson emission computed from eqn.\splitX, while ${\cal P}^g$
is the splitting function due to gluon
emission.

Here, we are interested in studying the effects of the
generation of an asymmetric sea due to the nonperturbative splitting
function  ${\cal P}^\Pi$; however, we need a model of the gluon
emission mechanism in order to be able to obtain predictions which may
be compared with the data. The perturbative gluon splitting function
is itself inadequate because we wish to begin our evolution at a low
scale $Q^2\sim 0.04$~GeV$^2$ where the nonperturbative evolution
due to meson emission flattens \balfor\ and the sea and gluon
distributions may be reasonably
assumed to vanish; at such low scales perturbation theory is clearly
useless. Indeed, early studies~\ref\radpart
{G.~Parisi and R.~Petronzio, \PL\vpy{B62}{331}{1976}\semi
V.A.~Novikov, M.A.~Shifman, A.I.~Vainshtein and V.I.~Zakharov,
\AP\vpy{105}{276}{1977}\semi
M.~Gl\"uck and E.~Reya, \NP\vpy{B130}{76}{1977}\semi
M.~Gl\"uck, R.M.~Godbole and E.~Reya, \ZP\vyp{C41}{1989}{667}.}
show that at large $Q^2\gsim$~1~GeV$^2$ the perturbatively generated
sea and gluon distributions rise far too steeply at small $x$. More
reasonable distributions may be obtained~\ref\niko{V.~Barone \etal,
\ZP\vpy{C58}{541}{1993}; \IJMP\vpy{A8}{2779}{1993}.} by suitably
suppressing the emission of very soft
gluons, and this is the approach we will follow here.\foot{An
alternative possibility~\ref\glre{M.~Gl\"uck,
E.~Reya and A.~Vogt, \ZP\vpy{C48}{471}{1990}; \vpy{C53}{127}{1992}.} is to
assume a large gluon component
with a valence--like distribution at very low scales (``valence
gluons''). We will not consider this option as it would require us to
introduce for consistency also an ansatz for the starting sea
asymmetry; the connection with the constituent quark model would then
be lost, and the evolved distributions would become almost entirely
model dependent.}

We will thus model the evolution of the valence distribution by using the
generalized splitting function of ref.\niko. This is computed by
freezing the strong coupling below $Q=0.5$~GeV, and then assuming
instead that the gluons have a `mass' $\mu_G$ of several hundreds of
MeV; this ensures that the evolved gluon distributions will be harder
than the corresponding perturbative ones. The flux of gluons generated
by the target quark is then determined by solving an appropriate
evolution equation which factorizes into an Altarelli--Parisi--like equation
only at scales rather larger that the gluon mass.
Gluon and sea quark distributions
are thus generated dynamically by evolving
a valence quark distribution.
The starting valence distribution can be
 obtained from a three--quark
 wave function defined in the limit of small $Q^2$;
it has been shown \niko\ that a simple gaussian wave function,
whose width is fixed in terms of  the r.m.s. radius
of the proton, yields
both a correct shape for the gluon distribution and a
correct balance between valence and sea components.
The gaussian ansatz is sufficient for
our purposes, since we are only really interested in the scale dependence of
the sea asymmetry, which is due almost entirely to the nonperturbative
evolution induced by meson emission; we will check that our results do not
depend on the detailed form of the starting valence evolution by varying the
shape of this ansatz.

We have therefore solved the evolution equation \evolqp,\qdef\ with the
splitting functions eqn.\splf,  ${\cal  P}^\Pi$ from ref.\balfor, and
${\cal P}^g$ from ref.\niko. Because the nonperturbative coupling to
bound states generates a  contribution to the nonsinglet sea $\bar
u-\bar d$, it will also contribute to the singlet structure function:
clearly $|\bar u +\bar d|\ge|\bar u -\bar d|$. But then
the parameters which control
the valence evolution according to the splitting function ${\cal P}^g$
must be adjusted in order to insure that the balance of the fractions
of nucleon momentum carried by valence, sea and gluons is kept in
agreement with experiment; the unobserved
fraction of the total momentum (namely, the momentum which is not
carried by quarks or antiquarks) should
be around 45\% at 4~GeV$^2$. Due to the nonperturbative
mechanism, an unobserved momentum fraction carried by bound states
is  generated, and the momentum carried by the gluons must be
accordingly decreased. This is done by computing
the second moment of the nonperturbatively generated
nonsinglet quark--antiquark
sea, using it to estimate the momentum that goes into the bound
states,
and then readjusting $\mu_G$ in order to reduce the momentum
fraction carried by the gluon sea by the same amount.

With
typical values of the parameters (see below) the momentum fraction
carried by the nonperturbatively generated sea at $Q^2=0.5$~GeV$^2$ is
around 15--20\% of the total momentum, the momentum carried by the
glue is 20--25\%, and the valence and perturbatively generated sea quarks
carry the rest. As $Q^2$ grows further, the nonperturbative sea generation
flattens rapidly and eventually disappears above $Q^2\sim$~5~GeV$^2$; standard
perturbative evolution, and the usual balance of momentum
between quarks and gluons, are then  recovered.

{}Firstly, we consider the scale dependence of the full nonperturbative
structure function, displayed in \fig\sfevol{The structure function
$F_2^{p}-F_2^{n}$ for various values of $Q^2$. }.
The evolution starts at a
very low scale $Q^2=0.04$~GeV$^2$, where the nonperturbative evolution
flattens \balfor\ and it can thus be assumed that there is no sea asymmetry.
The valence distribution is constructed from a gaussian nucleon wave
function~\niko, whose radius is determined with an uncertainty of
the order of $20 \%$. The values of
the two parameters which control the nonperturbative evolution
are fixed by requiring the value of the
Gottfried sum rule measured experimentally~\refs{\nmca,\nmcb}
\nref\shad{N.N.~Nikolaev and B.G.~Zakharov, \ZP\vyp{C49}{1991}{607};
\PL\vyp{B260}{1991}{414}\semi
V.~R.~Zoller, \PL\vpy{B279}{145}{1992}\semi
B.~Bade\l ek and J.~Kwieci\'nski, \NP\vpy{B370}{278}{1992}\semi
V.~Barone \etal, \ZP\vyp{C58}{1993}{541};
\PL\vpy{B321}{137}{1994}\semi
 W.~Melnitchouk and A.W.~Thomas, \PR\vyp{D47}{1993}{3783}.}
at $Q^2=$~4~GeV$^2$ to be correctly reproduced.\foot{Notice that
the value published originally \nmca\ $S_G=0.24\pm
0.016$ was recently reevaluated \nmcb\ to give $S_G=0.258\pm0.017$. However,
the data which lead to this reevaluation also show evidence for shadowing
effects in deuterium, which were neglected in both analyses. Taking
current theoretical estimates \shad\ for these effects and correcting the new
data leads back to a value  $S_G\approx 0.20-0.24$.}
Combined with the theoretical expectation that $g_\pi\lsim {1\over2}$,
this fixes
\balfor\ $\Lambda=0.4$~GeV
with an uncertainty  of order
20\%.\foot{The dependence of the results on $g_\pi$ is actually rather
weak, because
 $g_\pi$ only affects  the large $Q^2$ tail of the
evolution of the first moment (which is weakly evolving),
hence the asymptotic value of $S_G$,
rather than the measured one.}
 As may be seen from the figure, the scale
dependence of the nonsinglet structure function turns out
to be rather strong in the medium--large $x$ region, $0.15<x<0.6$.

The nonsinglet structure function at $Q^2=4$~GeV$^2$ is then compared in
\fig\sfexp{Comparison of the calculated nonsinglet structure function
at $Q^2=4$~GeV$^2$ (solid curve of \sfevol) with the
data of ref.\nmcb. The dotted curve displays the result obtained with
${\cal P}^{\Pi}=0$ in eqn.\splf\ and the dot-dashed curve the result obtained
setting ${\cal P}^g=0$.}
with the most recent data from the NMC\nmcb \foot{
An alternative determination of
the nonsinglet distribution, which is peaked at lower $x$ ($x\sim 0.35$),
has been offered in
ref.\ref\kabu{E.M.~Kabu\ss~(NMC),
 {\it Nucl. Phys. B (Proc. Suppl.)} \vyp{29A}{1992}{1}\semi
A.~Br{\"u}ll~(NMC), Ph.D. Thesis, Freiburg University, 1993.}, but
never published. Our computation is in fair agreement
with  this analysis too.}.
The figure also shows the nonsinglet structure
function obtained from the model of ref.\niko, \ie, with the
evolution due to meson emission switched off (${\cal P}_\Pi\equiv0$),
and thus with a flavour symmetric sea.
Comparing this curve with the data demonstrates a problem common to most
dynamical evolution mechanisms: because initially the valence quarks
carry all of the nucleon's momentum, the nonsinglet distribution is typically
peaked at $x\approx\third$; if the evolution of this distribution is
entirely due to the emission of flavour singlet objects (for example
gluons) then since the momentum of the valence quarks is
necessarily degraded, at large scales the maximum of the distribution
occurs at a  lower value, $x\sim 0.1-0.2$.\foot{Indeed the
valence distributions extracted from fits to structure function
data~\ref\MRS{A.~D.~Martin, W.~J.~Stirling and R.~G.~Roberts,
{\it J. Phys.} \vyp {G19}{1993}{1429}.}
appear to have a maximum located at $x\sim 0.2$.}
This disagrees with the data \nmcb,\kabu, which are instead peaked
at $x\approx 0.35-0.45$.

By contrast, the nonperturbative evolution generated by emission of
bound states leads to a net increase in the number of sea quark
pairs; the charge conjugation even first moment is no longer conserved
(even though the charge conjugation odd is, of course, in agreement
with the Adler sum rule). It is then possible for the sea
distributions to become
harder, rather than softer, and indeed it turns out that this is the
case: if the nonperturbative
mechanism were the only one responsible for evolution,
the maximum of the distribution would
shift to larger, rather than smaller values of $x$ (compare the
dash-dot curve in \sfexp).
Because the flavour off--diagonal nonperturbative emission is favoured with
respect to the flavour diagonal one \balfor, the  $\bar d$ distribution
in eqn.\nsstrf\  increases more rapidly than the $\bar u$ one, and the
net effect of the nonperturbative evolution is to make the full
nonsinglet distribution eqn.\nsstrf\ decrease more rapidly than its
valence component. This effect is sufficient to bring the calculated
 distribution in line with the data\nmcb,
because it  counteracts the degrading effect of the gluon emission.
The nonsinglet nonperturbative evolution mechanism is thus probably
the key ingredient in the determination of the shape of the
nonsinglet structure function.

The sensitivity of this result to the choice of parameters of the
nonperturbative evolution is tested in \fig\sfpar{Dependence of the
nonsinglet structure function  on the parameters
$\Lambda$ and $g_\pi$. The NMC data are also shown.},
where the variation of
the structure function at $Q^2=4$~GeV$^2$ is shown as $\Lambda$ and
$g_\pi$ are varied within their admissible ranges discussed above.
It appears that the sensitivity to $g_\pi$ (which
cannot be fixed accurately from independent data) is
minimal, and concentrated in the small $x$ region; the sensitivity to
$\Lambda$
is rather more significant. The
dependence of the results on the precise form of the ansatz for the
starting valence distribution is shown in
\fig\sfval{Comparison of the nonsinglet structure functions obtained from
two different
initial valence distributions: the gaussian wave function valence
distribution of ref.\niko\ (dot-dashed curve at the input scale,
solid curve at $Q^2 = 4$~GeV$^2$) and the power--like valence
eqn.\startv\
(dotted curve at the input scale, dashed curve
at $Q^2 = 4$~ GeV$^2$). The NMC data are also shown.},
which compares to the previous results those  obtained
choosing a power like form for the starting valence,
\eqn\startv
{u^v(x)-d^v(x)=C \, x^{0.5} (1-x)^{2}\quad\hbox {at 0.04 GeV}^2,}
where the values of the exponents are determined from the observed
 valence distribution.
It appears that indeed
the $Q^2$ evolution of the nonsinglet structure function
is largely independent of the precise
form of the valence distribution which is chosen.

 It follows from this discussion that a measurement of the
$Q^2$ dependence of the nonsinglet nucleon structure function
in the range 1~GeV~$^2\lsim Q^2\lsim 5$~GeV$^2$  allows an experimental test
of the nonperturbative sea generation mechanism.
A detailed set of predictions is thus provided in
\fig\sfqs{The $Q^2$ dependence of the structure function at various
values of $x$ (solid curves) compared with the standard $Q^2$ dependence
obtained
omitting the nonperturbative evolution mechanism (dashed lines). Note
the change of vertical scale between frames.},
where the $Q^2$ dependence of the structure
function over a suitable range of $Q^2$ is shown and compared with the scale
dependence due to gluon emission alone (applied to the same purely
valence starting distribution at $Q^2=0.04$~GeV$^2$).
It is apparent that the nonperturbative nonsinglet evolution is
maximal (and differs most strongly from the purely gluonic evolution) in
the central region of $Q^2$ indicated above, and in the region
$0.1\lsim x \lsim 0.35$ where the maxima of the two evolved distributions
are located. For larger values of $Q^2$ the perturbative evolution
is regained; since the evolution is multiplicative,
this means that, as a function of $Q^2$,
the structure function generated through nonsinglet
evolution becomes eventually  parallel (rather than equal) to that evolved
by gluon emission, as displayed in \sfqs\ for $Q^2\gsim 5$~GeV$^2$.

Figs.\xfig\sfexp\ and \xfig\sfqs\ also  show that the effect
which is responsible for violation of the Gottfried
sum rule \balfor, rather than being concentrated at small $x$, is
relevant in the intermediate $x$ region as well.
Specifically,
 computing the difference $\Delta S_G$
between the value of the Gottfried sum $S_G$ (\ie, the first moment of the
nonsinglet structure function eqn.\nsstrf) which we get  at $Q^2=4$~GeV$^2$,
and the naive parton model
prediction $S_G=1/3$,
we find that only about $55 \%$ of $\Delta S_G$ comes from the region
$x< 0.05$;
$20 \%$ comes from  the region $0.05 \le x \le 0.1$,
and $25 \%$ comes from the region $0.1 \le x \le 0.4$.
Therefore, the nonperturbative effect studied here
significantly affects the scale
dependence of higher moments of the nonsinglet structure function, and
not only the first moment.

A flavour asymmetry in the quark--antiquark sea may also
be measured directly in
Drell--Yan production, which allows~\ref\DYth{S.D.~Ellis
and W.J.~Stirling, \PL\vpy{B256}{258}{1991}.}
a determination  of the asymmetry ratio
\eqn\ratdef
{R\equiv{\bar d(x)-\bar u(x)\over \bar d(x)+\bar u(x)}}
from the comparison of
the cross sections for production on isoscalar targets
with that on targets with a proton or neutron excess. We can provide a
prediction for this quantity by extracting the sea asymmetry in the
numerator from the nonsinglet structure function according to the
decomposition eqn.\nsstrf, where the valence distribution $q^v$ at any
scale is
determined as the nonsinglet structure function obtained by setting the
nonperturbative
splitting function  ${\cal P}^\Pi=0$ in the evolution equations. The
denominator is then found by computing the symmetric portion of the
sea according to the procedure of ref.\niko\ (but with the parameters
of the gluon distribution suitably adjusted to ensure that the
momentum sum rule is correctly satisfied in the presence of an
asymmetric sea, as discussed above)  and adding to it the
asymmetric sea determined as above.  We use this self-consistent
procedure,
rather than, for example, computing the numerator and taking the
denominator from available fits to the data, in order to minimize the
error due to the very poor experimental knowledge of the sea
distributions for $x\gsim 0.15$, where the sea is tiny
and drops very rapidly with increasing $x$.

Our prediction for the ratio
$R$ , eqn.\ratdef, is shown in
\nref\na{NA51 collaboration; P.~Sonderegger \etal,
proposal CERN SPSLC/92-15/P267 (1992);
and work in preparation, B.~Alessandro,
and M.~Monteno (NA51), private communication.}
\nref\DYa{P.~L.~McGaughey \etal (E772), \PRL\vpy
{69}{1726}{1992}.}\nref\kl{S.~Kumano and
J.~T.~Londergan, \PR\vpy{D44}{717}{1991}.}
\nref\ehq{E.~J.~Eichten, I.~Hinchliffe and C.~Quigg, \PR\vpy{D45}{2269}
{1992}.}\fig\dy{Our computation of the
asymmetry ratio $R$ eqn.\ratdef\
compared with the available
experimental information from the
Drell--Yan process. The curve  is given at $Q^2= 10$~ GeV$^2$,
though it  is essentially $Q^2$--independent
for $Q^2\gsim 5$~GeV$^2$.
The square is the NA51 determination \na ($Q^2=25$~GeV$^2$). The
region
below the finely dotted curve
is the area allowed by the E772 data ($25\lsim Q^2 \lsim 170$~GeV$^2$)
at a $2 \sigma$ statistical
error level \DYa. The broadly dotted, dot-dashed, and dashed curves
are predictions from the models of refs.\kl\ and \ehq, and the fit of
ref.\DYth,
respectively (see ref.\DYa).}.
At present, the most precise determination of $R$ (although
limited to one point in the  $(x,Q^2)$ plane, namely, $x=0.18$,
$Q^2 \simeq 25$~GeV$^2$) comes from the NA51
experiment on Drell--Yan production on deuteron \na;
there are also upper limits on $R$ in a broad $x$ range, obtained
from Drell--Yan production
on tungsten \DYa.
The data are in good agreement
with our calculation. Of course one should always keep in mind
that there is an intrinsic theoretical uncertainty in our
determination of $R$,  due
to the need to estimate the singlet structure function which
contributes to the denominator of eqn.\ratdef. However, this does not
affect the overall trend of our prediction (specifically, the fact
that as a function of $x$ $R$ should flatten for $x\gsim 0.2$)
and  data
 in a broader $x$-range should provide a test of it.

Finally, it is interesting to see how  a
nonperturbative contribution to the scale dependence of the nonsinglet
structure function would affect
the  scale
dependence of the ratio
 $F_{n/p} \equiv F_2^n/F_2^p$. Indeed, there exists an indication that
the observed scale dependence of $F_{n/p}$ disagrees with the prediction of
perturbative QCD~\ref\nmcc{P.~Amaudruz \etal (NMC),
\NP\vyp{B371}{1992}{3}.}. Even though
the experimental situation is  inconclusive, due to the large
errors involved in the determination of the scale dependence, it would appear
that $\gamma_{n/p}(x)={d\over dt}F_{n/p}(x)$ is
negative
at small $x\lsim 0.3$ and positive at large $x\gsim 0.4$,
whereas numerical solution
of the (perturbative) Altarelli--Parisi equations  gives~\ref\virch{M.~Virchaux
and A.~Milsztajn, \PL\vyp{B274}{1992}{221}.} negative values
of $\gamma_{n/p}$ up to
$x\sim 0.6$; furthermore at small $x$ the measured value of
$|\gamma_{n/p}|$ appears to be larger by perhaps one order of
magnitude than that expected perturbatively.

A precise determination of $\gamma_{n/p}$ requires a computation of
the scale dependence of the singlet structure functions, and thus
goes beyond the scope of
the present work, where we concentrate on nonsinglet evolution.
However, using the fact that the nonperturbatively generated asymmetry
satisfies \hbox{$|{d\over dt} (\bar u-\bar d)^{NP}|\le
|{d\over dt} (\bar u+\bar d)^{NP}|$}
it is easy to see that the nonperturbative contribution to $\gamma_{n/p}$
has the opposite sign to \hbox{${d\over dt}(F_2^p-F_2^n)$.}
This means that (compare \sfqs)
the effect discussed here will lead to a
negative contribution to $\gamma_{n/p}$ for $x\lsim 0.1$ and a
positive one  for $x\gsim 0.15$ growing quite large around $x\sim
0.25$, in qualitative agreement with the experimental trend. A more
conclusive study of this issue will require an improvement in the
accuracy of the experiments, as well as a better theoretical
understanding of the singlet evolution.

\medskip

In conclusion, we have discussed the consequences for the scale
dependence of the nonsinglet nucleon structure function of the
nonperturbative generation of a sea quark flavour asymmetry, due
essentially to the anomalous breaking of the U(1) symmetry in the
spectrum of pseudoscalar mesons. These contribute to the generalized
Altarelli--Parisi equations constructed in ref.\balfor, through
splitting functions for the emission of bound states from quarks.
The nonperturbative evolution, besides explaining \balfor\ the
observed~\refs{\nmca,\nmcb}\ violation of the Gottfried sum rule as a
consequence of strong evolution of the first nonsinglet moment in the
region $0.5 \lsim Q^2\lsim 5$~GeV$^2$, predicts a stronger scale dependence
of the full structure function in the same $Q^2$ range for medium--large
values of $x$, $0.1\lsim x\lsim 0.6$. This is due to the fact that the
nonperturbative symmetry breaking allows for the generation of
a relatively hard sea component, peaked around $x\approx 0.35$, which,
due to its flavour asymmetry, contributes to the nonsinglet structure function.
Perturbative evolution is regained at large $Q^2\gsim$5~GeV$^2$.

The nonperturbatively generated sea asymmetry leads to a considerable
improvement in the agreement between the computed nonsinglet structure
function and available data, quite independently of the way
the initial valence distribution is modeled and evolved, thus suggesting that
indeed the model presently under discussion is responsible for the
bulk of the generation of the nonsinglet sea. The model can
be tested by measuring the scale dependence of the nonsinglet
structure function, for which we have presented a detailed set of
predictions. We have also provided a
prediction for the sea asymmetry as measured in Drell--Yan production
experiments; our prediction is compatible with available data which
however do not allow any definite conclusion.
If confirmed, the effects discussed here and in
ref.\balfor\ would provide the first conclusive evidence for
scale dependence which cannot be obtained through
purely perturbative techniques.
\medskip
{\bf Acknowledgements:} We thank G.~Altarelli, M.~Anselmino,
E.~Predazzi, R.~D.~Roberts  and the Torino component
of the NA51 Collaboration for discussions.

\vfill\eject
\listrefs\vfill\eject
\listfigs\vfill\eject
\bye